\documentclass[%
 aip,
 amsmath,amssymb,
 reprint,%
]{revtex4-1}

\usepackage{graphicx}% Include figure files
\usepackage{dcolumn}% Align table columns on decimal point
\usepackage{bm}% bold math
\usepackage[utf8]{inputenc}
\usepackage[T1]{fontenc}
\usepackage{mathptmx}
\usepackage{etoolbox}
\usepackage{xcolor}
\usepackage{ulem}
\usepackage{amsmath}
\usepackage{amssymb}

\makeatletter
\def\@email#1#2{%
 \endgroup
 \patchcmd{\titleblock@produce}
  {\frontmatter@RRAPformat}
  {\frontmatter@RRAPformat{\produce@RRAP{*#1\href{mailto:#2}{#2}}}\frontmatter@RRAPformat}
  {}{}
}%
\makeatother

\DeclareMathOperator{\sinc}{sinc}

\begin{document}

\preprint{AIP/123-QED}

\title{Slow spectral dynamics of shot noise in the Kuramoto model: the role of microscopic regularity}
% Force line breaks with \\
\author{S. Yu. Kirillov}
 \email{skirillov@ipfran.ru}
 \affiliation{A.V. Gaponov-Grekhov Institute of Applied Physics of the Russian Academy of Sciences, Ulyanova Street 46, Nizhny Novgorod 603950, Russia%\\This line break forced with \textbackslash\textbackslash
}%
\author{V. V. Klinshov}%
 %\altaffiliation[Also at ]{National Research University Higher School of Economics}%Lines break automatically or can be forced with \\
 \email{vladimir.klinshov@ipfran.ru}
  \affiliation{A.V. Gaponov-Grekhov Institute of Applied Physics of the Russian Academy of Sciences, Ulyanova Street 46, Nizhny Novgorod 603950, Russia%\\This line break forced with \textbackslash\textbackslash
}%
 \affiliation{National Research University Higher School of Economics, 25/12 Bol’shaya Pecherskaya Street, Nizhny Novgorod 603155, Russia%\\This line break forced with \textbackslash\textbackslash
}%

\begin{abstract}
Finite-size effects in the Kuramoto model are known to induce collective fluctuations even below the critical coupling, where the thermodynamic limit predicts complete asynchrony. While the shot-noise approach developed in our recent work accurately describes the power spectrum of these fluctuations for random frequency sampling, the present study reveals that the microscopic realization of the frequency distribution plays a crucial role. We show that a deterministic (quasi-uniform) selection of natural frequencies from the same Lorentzian distribution leads to qualitatively different dynamics: the shot noise spectrum exhibits anomalously slow oscillatory behavior, manifesting as wave-like patterns in time-frequency representations. The period of these oscillations scales linearly with the system size and matches the frequency spacing between neighboring oscillators near the distribution center. Numerical simulations confirm that these slow spectral dynamics arise from resonant interactions facilitated by the regular frequency structure, which are absent for random sampling. Our findings demonstrate that identical integral frequency distributions do not guarantee equivalent collective dynamics, highlighting the necessity of accounting for the fine structure of microscopic parameters in finite-size populations.
\end{abstract}

\maketitle

\begin{quotation}
The Kuramoto model is a classical paradigm of nonlinear dynamics, widely used to study synchronization in systems of coupled oscillators — from neurons firing in the brain to generators in power grids. While there is a well-established theory for infinitely large systems, real-world networks are finite, and their dynamics often deviate from mean-field predictions. In a recent work, we showed that below the critical coupling, finite-size effects induce collective fluctuations — a phenomenon we termed ``shot noise'' — and derived an analytical expression for their power spectrum. Here, we uncover a surprising twist: the way natural frequencies are chosen matters. When frequencies are selected deterministically rather than randomly, the system exhibits ultra-slow oscillations in its fluctuation spectrum, absent in the random case. The period of these oscillations scales with the system size, revealing that hidden regularity in the frequency distribution can dramatically alter collective dynamics. This finding underscores that identical statistical distributions do not guarantee identical behavior in finite systems, a crucial insight for understanding synchronization in real-world applications ranging from neural networks to power grids.

\end{quotation}

\section{Introduction}

Large-scale ensembles of coupled self-oscillatory elements serve as a fundamental model for describing a wide range of phenomena in physics, biology, and engineering — from brain neural networks \cite{Wilson2022,Breakspear2010,Maslennikov2015} to arrays of Josephson junctions \cite{Wiesenfeld2020} and power grids \cite{Filatrella2008,Khramenkov2026}. A key challenge in studying such systems is identifying patterns of their collective dynamics, which can often be described by a small number of macroscopic parameters, despite the high complexity of the microscopic dynamics of individual elements. A fundamental approach to reducing such systems was proposed by Winfree \cite{Winfree1967} and mathematically formalized by Kuramoto \cite{Kuramoto1984}, who showed that for weakly coupled oscillators, the system dynamics can be reduced to the evolution of their phases. The Kuramoto model and its generalizations \cite{Kuramoto1984,Shinomoto1986} have become a standard paradigm for studying synchronization phenomena. Further theoretical developments led to the creation of effective mean-field models \cite{Watanabe1994,Pikovsky2008,Sakaguchi1988,Daido1996,Klinshov2021,Tyulkina2018,Goldobin2021,Ageeva2025}, including the Ott-Antonsen approach \cite{Ott2008,Ott2009,Montbrio2015}, which excellently describe the behavior of systems in the thermodynamic limit where the number of elements tends to infinity.

However, real networks always have a finite size, raising the question of the role of finite-size effects, which can lead to qualitatively new phenomena absent in infinite systems \cite{Ageeva2024,Goldobin2024,Zhang2025,Goldobin2025}. Previously, we developed an approach based on the concept of ``shot noise'' \cite{Klinshov2021v2}, which allows describing the stochastic mean-field dynamics as a result of fluctuations due to the discreteness of elements. The proposed concept enables describing the properties of collective fluctuations in finite-size populations near the stationary solution of the infinite system through spectral characteristics. It was shown that shot noise is colored, can have pronounced peaks in the power spectrum, and can lead to phenomena not predicted within classical mean-field models \cite{Klinshov2023,Kirillov2023,Kirillov2024}.

In a recent work \cite{Kirillov2025}, we applied this approach to the classical Kuramoto model and showed that in the subcritical region ($K < 2$), where the thermodynamic limit predicts complete asynchrony, the finite size of the system induces collective fluctuations — shot noise. We obtained an analytical expression for the power spectrum of these fluctuations, investigated their variance and correlation time, and demonstrated good agreement between theoretical results and numerical simulations in the region away from the bifurcation point. It is important to note that in that work, as in the vast majority of studies on finite-size effects, the natural frequencies of the oscillators were chosen randomly from a given distribution. This approach, however, is not the only possible one: the element frequencies can also be set in a regular manner. The importance of accounting for the microscopic realization of the frequency distribution was demonstrated in the work by Hong et al. \cite{Hong2015}, where it was shown that random and quasi-uniform (regular) sampling of frequencies from the same distribution leads to significantly different scaling properties of the order parameter and its fluctuations near the synchronization transition point. In a recent study by Park and Park \cite{Park2024}, these differences were investigated further: the finite-size scaling properties for deterministic frequency sampling were found to differ substantially from previously known results for random sampling.

In the present work, we investigate the influence of the method of setting oscillator frequencies on the properties of shot noise in the Kuramoto model. We find that deterministic quasi-uniform frequency selection leads to the emergence of a qualitatively new phenomenon: ultra-slow oscillations appear in the shot noise spectrum, which are absent for random frequency realizations. This effect, presumably related to the fine ordered structure of the frequency distribution, can play a significant role in the dynamics of finite ensembles, leading to long transient processes and metastable states. This expands the understanding of finite-size effects beyond classical mean-field models.

The paper is organized as follows. Section 2 presents theoretical results on the shot noise spectrum in the finite-size Kuramoto model. Section 3 presents the results of numerical investigations of the Kuramoto model with both random and deterministic distributions of element frequencies. In Section 4, we discuss our results and draw brief conclusions.

\section{Theory}\label{sec:model}

As a basic model in this study, we consider the classical network of Kuramoto oscillators. This system is widely known as a paradigmatic example demonstrating a phase transition from an asynchronous state to collective synchronization as the coupling strength increases. In this work, the main focus is on the behavior of the finite-size system in the subcritical (asynchronous) regime. The finite number of natural frequencies in the ensemble gives rise to noise-like fluctuations of the complex order parameter — a quantity describing the collective rhythm of all oscillators in the network. We demonstrate that the spectral properties of these fluctuations can be qualitatively and quantitatively described within the framework of the nestling principle and the shot noise concept we previously proposed. In contrast to our previous work \cite{Kirillov2025}, the present study conducts an in-depth analysis of the influence of the internal (microscopic) structure of the natural frequency distribution on the collective (macroscopic) dynamics of the oscillator network. We find that the identity of the integral distribution does not guarantee uniformity of collective behavior: the presence of a partially regular structure within it initiates complex transient processes and causes a qualitative change in the spectrum of collective fluctuations not described by existing theoretical models. The analysis of the temporal evolution of the spectrum allows not only to identify the key physical mechanisms responsible for the observed deviations but also lays the foundation for constructing new generalized theoretical descriptions that account for the influence of explicit or hidden regular structures on collective dynamics.

The model has the form:
\begin{equation}\label{a1_01}
\frac{d\theta_i}{dt}=\omega_i+\frac{K}{N}\sum_{j=1}^N \sin(\theta_j-\theta_i),\quad i=1\dots N,
\end{equation}
where $\theta_i$ is the phase of the $i$-th oscillator ($\theta\in[0;2\pi)$), $\omega_i$ is its natural frequency, $K$ is the global coupling strength between all oscillators, and $N$ is the total number of elements in the network.

As an integral characteristic of the collective state of such a network, it is convenient to use the complex Kuramoto order parameter:
\begin{equation}\label{a1_02}
R=\frac{1}{N}\sum_{j=1}^N e^{i\theta_j}.
\end{equation}

Below, the mean-field approximation of model (\ref{a1_01}) is presented, and then the analysis is extended to the finite-dimensional case using the proposed approach \cite{Kirillov2025}. This will allow us to analytically describe the properties of collective fluctuations in the subcritical region ($K<2$).

\subsection{Asymptotic behavior of the network in the thermodynamic limit}

As the number of network elements tends to infinity ($N\rightarrow \infty$), an asymptotic transition occurs from the discrete microscopic description of the dynamics of individual oscillators to their macroscopic description through a continuous probability density $f(\omega, \theta, t)$ \cite{Kuramoto1987,Strogatz1991}. The quantity $f(\omega, \theta, t) d\omega d\theta$ in this case determines the fraction of oscillators at time $t$ with natural frequencies in the interval $[\omega, \omega + d\omega]$ and phases in the interval $[\theta, \theta + d\theta]$. The normalization condition for the total number of oscillators is
\begin{equation}\label{a2_01}
\int_{-\infty}^{\infty}\int_0^{2\pi}f(\omega,\theta,t)d\theta d\omega=1,
\end{equation}
and the integral over all phases
\begin{equation}\label{a2_02}
\int_0^{2\pi}f(\omega,\theta,t)d\theta=g(\omega),
\end{equation}
defines the time-stationary continuous distribution of natural frequencies $g(\omega)$.

The time evolution of the density $f$ obeys the conservation law for the number of oscillators, expressed by the following continuity equation:
\begin{equation}\label{a2_03}
\frac{\partial f}{\partial t}+\frac{\partial}{\partial\theta}\bigg\{\big[\omega+\frac{K}{2i}(r e^{-i\theta}-r^* e^{i\theta})\big]f\bigg\}=0,
\end{equation}
where the order parameter $r$ of the infinite-dimensional system is given by the integral expression:
\begin{equation}\label{a2_04}
r=\int_0^{2\pi}d\theta\int_{-\infty}^\infty d\omega f e^{i\theta}.
\end{equation}

Given the constraint $|r|\leq 1$ (which follows from definition (\ref{a1_02})), and the periodic nature of the phase variables $\theta$, the solution of equation (\ref{a2_03}) is represented as a Fourier series:
\begin{equation}\label{a2_05}
f=\frac{g(\omega)}{2\pi}\bigg\{1+\sum_{n=1}^\infty \big[ f_n(\omega,t)e^{i n\theta}+c.c.\big]\bigg\},
\end{equation}
where $c.c.$ denotes complex conjugate.

Following the Ott-Antonsen approach \cite{Ott2008,Ott2009}, we consider a special class of functions $f_n(\omega,t)$ such that
\begin{equation}\label{a2_06}
f_n(\omega, t)=[\alpha(\omega, t)]^n,
\end{equation}
where $\alpha(\omega, t)$ is a complex function satisfying $|\alpha(\omega, t)|\leq 1$, ensuring the convergence of the Fourier series.

Substituting the special form of the solution (\ref{a2_06}) into the continuity equation (\ref{a2_03}) allows us to transition from an infinite set of equations for coupled modes to a closed system of equations for the auxiliary function $\alpha(\omega, t)$:
\begin{equation}\label{a2_07}
\frac{\partial \alpha}{\partial t} + \frac{K}{2}\bigl(r \alpha^2 - r^*\bigr) + i\omega \alpha = 0,
\end{equation}
\begin{equation}\label{a2_08}
r(t)=\int_{-\infty}^{\infty}\alpha^*(\omega,t)g(\omega)d\omega,
\end{equation}
where $(\cdot)^*$ denotes complex conjugation.

We next consider the case where the distribution of natural frequencies of the oscillators is given by a Lorentzian function:
\begin{equation}\label{a2_09}
g(\omega)=\frac{1}{\pi}\frac{\Delta}{(\omega-\omega_0)^2+\Delta^2}.
\end{equation}

Given the rational form of distribution (\ref{a2_09}), the integral in (\ref{a2_08}) can be evaluated using the residue theorem. The integration contour must be closed in the lower half-plane, which ensures the physical realizability of the solution (\ref{a2_07}). The residue at the pole $\omega=\omega_0-i\Delta$ gives $r(t)=\alpha^*(\omega_0-i\Delta,t)$. Without loss of generality, we can normalize (\ref{a2_09}) by setting $\omega_0 = 0$ and $\Delta = 1$. Substituting $r(t)=\alpha^*(-i,t)$ and $\omega=-i$ into (\ref{a2_07}), we arrive at a system of equations for the amplitude $\rho$ and phase $\phi$ of the order parameter $r(t)=\rho(t) e^{i\phi(t)}$:
\begin{equation}\label{a2_10}
	\begin{aligned}
		&\frac{d\phi}{dt}=0,\\
		&\frac{d\rho}{dt}+\bigl(1-\tfrac{1}{2}K\bigr)\rho+\tfrac{1}{2}K\rho^3=0.
	\end{aligned}
\end{equation}

Analysis of system (\ref{a2_10}) shows that the amplitude equation has an equilibrium state $\rho=0$, corresponding to the asynchronous state of the full network (\ref{a1_01}). This state is stable (and unique) in the subcritical region $K<2$. At $K=2$, a supercritical Andronov-Hopf bifurcation occurs, leading to a loss of stability of the asynchronous state. In the supercritical region $K>2$, the stable equilibrium state is given by $\rho=\sqrt{(K-2)/K}$, corresponding to the transition of the system to the collective synchronization regime.

In \cite{Kirillov2025}, it was shown that unlike the asymptotic mean-field model (\ref{a2_10}), in systems of finite size, the asynchronous regime ($K<2$) is characterized by the presence of macroscopic fluctuations. The variance of these fluctuations $D=\langle\rho(t)^2\rangle-\langle\rho(t)\rangle^2$ (where $\langle\cdot\rangle$ denotes time averaging) is non-zero. Moreover, in the subcritical region, the transition to collective synchronization is preceded by an increase in the variance inversely proportional to the distance to the bifurcation point.

In the next section, we will demonstrate that the solution we obtained in \cite{Kirillov2025} for quantitatively describing the properties of collective fluctuations converges within the linear approximation at least over a finite interval of the subcritical region adjacent to the bifurcation point.

\subsection{Spectral properties of collective fluctuations in a finite-dimensional network}

Let us now analyze the case of a finite but sufficiently large oscillator network ($1\ll N<\infty$). Assume that the natural frequencies $\omega_i$ follow the previously introduced distribution function (\ref{a2_09}).

In the absence of coupling between elements ($K=0$), the phases of the oscillators (\ref{a1_01}) rotate uniformly:
\begin{equation}\label{a3_01}
\dot{\theta}_j=\omega_j.
\end{equation}

The order parameter (\ref{a1_02}), characterizing the collective dynamics of the finite-dimensional system, in this case takes the form:
\begin{equation}\label{a3_02}
R(t)=\frac{1}{N}\sum_{j=1}^N e^{i\omega_j t+i\theta_{0j}},
\end{equation}
where $\theta_{0j}$ denotes the phase of the $j$-th element at the initial moment.

Given the incommensurability of the natural frequencies and a sufficiently large number of elements $N$, the oscillations of the order parameter, while strictly speaking regular, acquire a noise-like character. Note that such noise-like oscillations exist regardless of the presence of interactions between elements, representing a direct manifestation of finite-size effects.

According to the results of our previous studies \cite{Klinshov2021v2,Klinshov2023,Kirillov2023,Kirillov2024}, despite the relatively small intensity of the noise (inversely proportional to the network size $N$), its influence on collective dynamics can be significant. In particular, it can lead to shifts in the boundaries of activity regimes, the formation of metastable states and switching dynamics, and the emergence of resonance effects. We will denote such noise-like oscillations in the system with mutual coupling as shot noise $\chi$, and in the absence of inter-element coupling as free shot noise $\chi_0$.

To analyze the properties of free shot noise $\chi_0$, we construct the autocorrelation function for the order parameter (\ref{a3_02}):
\begin{equation}\label{a3_03}
	\begin{aligned}
		C_{0}(\tau)&=\lim_{T\rightarrow\infty}\frac{1}{T}\int_0^T R(t)R^*(t-\tau)dt=\\
		&=\lim_{T\rightarrow\infty}\frac{1}{N^2 T}\int_0^T\sum_{j=1}^N e^{i\omega_j t+i\theta_{0j}}\sum_{k=1}^N e^{-i\omega_k t-i\theta_{0k}+i\omega_k\tau}dt=\\
        &=\lim_{T\rightarrow\infty}\frac{1}{N^2 T}\int_0^T\sum_{j=1}^N\sum_{k=1}^N e^{i(\omega_j-\omega_k)t+i(\theta_{0j}-\theta_{0k})+i\omega_k\tau}dt.
	\end{aligned}
\end{equation}

Analysis of (\ref{a3_03}) shows that the main contribution to the autocorrelation function comes from terms with $j=k$. The other terms represent periodic oscillations with amplitude $\sim 1/T$, and their contribution for large averaging times is negligible. Then,
\begin{equation}\label{a3_04}
C_{0}(\tau)=\lim_{T\rightarrow\infty}\frac{1}{N^2 T}\int_0^T\sum_{j=1}^N e^{i\omega_j\tau}dt=\frac{1}{N^2}\sum_{j=1}^N e^{i\omega_j\tau}.
\end{equation}

Based on this simplification, we find an analytical expression for the power spectral density of free shot noise:
\begin{equation}\label{a3_05}
    \begin{aligned}
		W_{0}(\omega)&=\int_{-\infty}^{\infty}C_{0}(\tau)e^{-i\omega\tau}d\tau=\int_{-\infty}^{\infty}\frac{1}{N^2}\sum_{j=1}^N e^{-i(\omega-\omega_j)\tau}d\tau=\\
        &=\frac{2\pi}{N^2}\sum_{j=1}^N\delta(\omega-\omega_j)\approx\frac{2\pi}{N}g(\omega).
	\end{aligned}
\end{equation}
The obtained expression demonstrates that, up to a constant factor inversely proportional to the number of network elements, the power spectral density of free shot noise coincides with the distribution function of the oscillator natural frequencies.

\begin{figure*}
\center{\includegraphics[width=14cm]{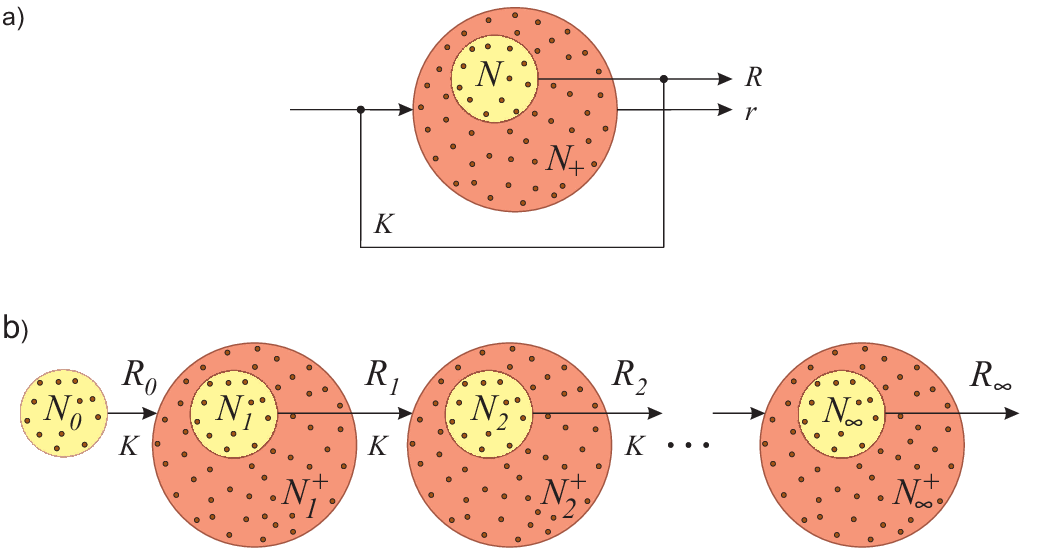}}
\caption{\textit{a} - Schematic of the nestling principle: the signal $R$ from the small network $N$ is fed to the input of the large network $N^+$. \textit{b} - Cascade of unidirectionally coupled identical ``nests''.} \label{fig:Nest_Kaskad}
\end{figure*}

Let us now analyze the influence of inter-element coupling on the collective dynamics of the system. To do this, we construct an auxiliary network using the previously proposed ``nestling principle'' \cite{Klinshov2021v2}. According to this approach, the original finite-size network of size $N$ is considered as a subset of a larger network $N^+$, whose number of elements is assumed to be unlimited ($N^+\rightarrow\infty$). In accordance with (\ref{a1_02}) and (\ref{a2_04}), we denote the output signal of the small (original) network as $R$, and the output signal of the large network as $r$.
The peculiarity of this approach is that the signal $R$ from the output of the small network is fed to the input of the large network (see Fig.~\ref{fig:Nest_Kaskad}, \textit{a}).
It is important to note that the operating conditions of the small network remain equivalent to the original problem.

Within the described configuration, the free shot noise $\chi_0$ in the small finite-size network $N$ is naturally defined as the difference between the collective signal $R$ at the output of this network and the collective signal $r$ at the output of the large network:
\begin{equation}\label{a4_01}
\chi_0=R-r.
\end{equation}
In other words, the free shot noise is the difference between the outputs of the finite and infinite populations, given the same input \cite{Kirillov2025}.

The signal $R$, acting on the large network $N^+$, induces macroscopic fluctuations in it, which we will denote as $\psi$. Consequently, the signal $r$ can be decomposed into two components:
\begin{equation}\label{a4_02}
r=r_{\infty}+\psi,
\end{equation}
where $r_{\infty}$ is the response of the network $N$ in the limiting case $N\equiv N^+$ (or equivalently, the mean field of the original network as $N\rightarrow\infty$).

The shot noise $\chi$ in the network $N$ is then defined as
\begin{equation}\label{a4_02_v2}
\chi=R-r_{\infty}.
\end{equation}

The form of the signal $r_{\infty}$ can be found using mean-field theory (\ref{a2_10}). As shown earlier, in the subcritical region $r_{\infty}=0$. To determine the properties of the output signal $R(t)$, as well as the characteristics of the shot noise $\chi(t)$ and macroscopic fluctuations $\psi(t)$ in this region, we establish a relationship between these signals and the properties of the previously found free shot noise $\chi_0$ (\ref{a3_05}). To do this, we construct an iterative procedure based on a new structure in the form of a cascade of unidirectionally coupled identical ``nests'', schematically shown in Fig.~\ref{fig:Nest_Kaskad}, \textit{b}.
The initial element, corresponding to the zero level of this cascade, is a finite-size network $N_0$ whose elements do not interact with each other. The output signal of such a network, denoted $R_0$, represents free shot noise. Next, a recursive process is implemented, where the signal $R_k$ from the output of the $k$-th level of the cascade is fed to an infinite-dimensional network $N_{k+1}^+$, whose elements do not interact with each other. Each such network, in turn, contains an embedded finite-dimensional network $N_{k+1}$, topologically equivalent to the original network $N_0$ (i.e., possessing an identical structure of connections and distribution of natural frequencies). The output signal of this finite-dimensional network is denoted $R_{k+1}$, and the iterative process repeats for the next level of the cascade.

We will show that there exists an interval in the subcritical region, adjacent to the bifurcation point, over which the sequence of signals $R_k(t)$ converges in the linear approximation to some limiting solution $R_{\infty}(t)$:
\begin{equation}\label{a4_03}
\lim_{k\rightarrow\infty} R_{k+1}(t)=R_k(t)=R_{\infty}(t).
\end{equation}
The fulfillment of the convergence condition (\ref{a4_03}) means that in the limit $k\rightarrow\infty$, the signal at the output of the cascade (see Fig.~\ref{fig:Nest_Kaskad}, \textit{b}) becomes indistinguishable from the signal at the output of the finite-dimensional network with couplings (see Fig.~\ref{fig:Nest_Kaskad}, \textit{a}), i.e., the desired signal can be found as $R(t)= R_{\infty}(t)$.

Let us show that the proposed approach is applicable for analyzing the spectral properties of shot noise in the system with couplings. Consider the $k$-th element of the cascade ($k\geq 1$). The infinite-dimensional network $N_k^+$ is acted upon by the signal $R_{k-1}$ from the previous element. The state of the network $N_k^+$ is described by the probability density $f_k(\omega,\theta,t)$, which obeys the following continuity equation:
\begin{equation}\label{a4_04}
\frac{\partial f_k}{\partial t}+\frac{\partial}{\partial\theta}\Bigg\{\bigg[\omega+\frac{K}{2i}\Big(R_{k-1}e^{-i\theta}-R^*_{k-1}e^{i\theta}\Big)\bigg]f_k\Bigg\}=0.
\end{equation}
Following the previously applied Ott-Antonsen ansatz, we represent the solution of (\ref{a4_04}) as a Fourier series (\ref{a2_05}) within the class of functions (\ref{a2_06}). This leads, analogously to (\ref{a2_07}) and (\ref{a2_08}), to a closed system of equations:
\begin{equation}\label{a4_05}
\frac{\partial\alpha_k}{\partial t}+\frac{K}{2}(R_{k-1}\alpha_k^2-R_{k-1}^*)+i\omega\alpha_k=0,
\end{equation}
\begin{equation}\label{a4_06}
r_k^*=\int_{-\infty}^{\infty}\alpha_k(\omega,t)g(\omega)d\omega=\alpha_k(-i,t).
\end{equation}
We restrict further consideration to the subcritical region $K<2$. Given the small intensity of the shot noise ($\sim 1/N$), one can specify a neighborhood $\varepsilon$ of the bifurcation point $K=2$ such that for $K<2-\varepsilon$ the nonlinear terms in equation (\ref{a4_05}) can be neglected. The value of $\varepsilon$ depends on several factors, including the network size, accuracy requirements, and structural features of the system. However, in very large networks ($N\rightarrow\infty$), the intensity of collective fluctuations becomes vanishingly small, and the region of nonlinear influence shrinks to the bifurcation point ($\varepsilon\rightarrow 0$). This allows us to linearize equation (\ref{a4_05}) and use it to describe the system across the entire subcritical region:
\begin{equation}\label{a4_07}
\frac{\partial\alpha_k}{\partial t}-\frac{K}{2}R^*_{k-1}+i\omega\alpha_k=0.
\end{equation}
Substituting (\ref{a4_06}) into (\ref{a4_07}), we obtain the equation for the order parameter
\begin{equation}\label{a4_08}
\frac{d r_k}{d t}+r_k-\frac{K}{2}R_{k-1}=0.
\end{equation}

Considering that throughout the considered range of $K$ values, $r_{\infty}=0$, by analogy with (\ref{a4_02}) we obtain $\psi_k=r_k$ (here $\psi_k$ are the macroscopic fluctuations caused by external rather than internal noise).

To determine the spectral properties of the macroscopic fluctuations $\psi_k$ arising in the $k$-th cascade level under the influence of the signal $R_{k-1}$, we find the complex transfer function $S(\omega)$ of system (\ref{a4_08}):
\begin{equation}\label{a4_09}
S(\omega)=\frac{K}{2}\frac{1}{1+i\omega}.
\end{equation}
Then, the power spectral density of the macroscopic fluctuations, as well as the signal at the output of the first cascade level ($k=1$), are given by:
\begin{equation}\label{a4_10}
\psi_1=S(\omega)R_0,
\end{equation}
\begin{equation}\label{a4_11}
R_1=R_0+\psi_1=\Big[1+S(\omega)\Big]R_0.
\end{equation}
For the second cascade element ($k=2$), we correspondingly obtain:
\begin{equation}\label{a4_12}
\psi_2=S(\omega)R_1=S(\omega)\Big[1+S(\omega)\Big]R_0,
\end{equation}
\begin{equation}\label{a4_13}
R_2=R_0+\psi_2=\Big[1+S(\omega)+S(\omega)^2\Big]R_0.
\end{equation}
Continuing the recurrence sequence similarly, we find that the expressions for $\psi_k$ and $R_k$ at the $k$-th cascade level represent power series, which can be written in compact form:
\begin{equation}\label{a4_14}
\psi_k=\bigg[\frac{1-S(\omega)^k}{1-S(\omega)}-1\bigg]R_0,
\end{equation}
\begin{equation}\label{a4_15}
R_k=\frac{1-S(\omega)^k}{1-S(\omega)} R_0.
\end{equation}
Analysis of functions (\ref{a4_14}) and (\ref{a4_15}) shows that the corresponding series converge if finite limits $\psi_{\infty}=\lim_{k\rightarrow\infty}\psi_k$ and $R_{\infty}=\lim_{k\rightarrow\infty}R_k$ exist. For this, the condition
\begin{equation}\label{a4_16}
\lim_{k\rightarrow\infty} S(\omega)^k =0
\end{equation}
must be satisfied. This condition, in turn, holds when $\big|S(\omega)\big|<1$, which with (\ref{a4_09}) gives
\begin{equation}\label{a4_17}
\frac{\pm K}{2\sqrt{1+\omega^2}}<1,
\end{equation}
where the ``+'' sign corresponds to positive values of $K$, and the ``-'' sign to negative ones.
Given that inequality (\ref{a4_17}) must hold for all real $\omega$, and solving it for $K$, we obtain
\begin{equation}\label{a4_18}
-2<K<2.
\end{equation}
This condition ensures that as $k\rightarrow\infty$, the sequences of functions $\psi_k(t)$ and $R_k(t)$ (see (\ref{a4_14}) and (\ref{a4_15})) converge uniformly to the limiting functions
\begin{equation}\label{a4_19}
\psi_\infty=\frac{S(\omega)}{1-S(\omega)}R_0,
\end{equation}
\begin{equation}\label{a4_20}
R_{\infty}=\frac{1}{1-S(\omega)}R_0.
\end{equation}

Note that within the linear approximation in the range of $K$ values from (\ref{a4_18}), the output signal of the finite-dimensional network coincides with its shot noise $\chi(t)=R(t)=R_{\infty}(t)$. As a result, for this region we obtain the shot noise spectrum depending on the coupling strength:
\begin{equation}\label{a4_21}
W_{\chi}(\omega)=\bigg|\frac{1}{1-S(\omega)}\bigg|^2 W_0(\omega)=\frac{2\pi}{N}\bigg|\frac{1+i\omega}{1+i\omega-\frac{K}{2}}\bigg|^2 g(\omega).
\end{equation}
The obtained expression shows that in the subcritical region, as the coupling strength increases, the intensity of collective fluctuations at frequencies near the mean increases, which is a precursor to the transition to the collective synchronization regime.

\section{Numerical simulation results}

Let us now compare the obtained analytical results with the results of numerical simulations of system \eqref{a1_01}. The study showed that the characteristics of shot noise are determined not only by the integral form of the natural frequency distribution but also by their internal structure. To investigate this effect, two different methods of selecting $\omega_j$ values were considered:
\begin{itemize}
    \item random method, where the frequency of each element is generated independently according to the given distribution \eqref{a2_09} using the inverse transform method:
    \begin{equation}\label{a6_01}
        \omega_j=\omega_0-\Delta\tan \bigg(\frac{\pi}{2}\frac{N-2 n_j+1}{N+1}\bigg),
    \end{equation}
    where $n_j$ is the $j$-th element in the ascending sequence of $N$ random real numbers uniformly distributed on the interval $[1,N]$.
    \item deterministic method, where the frequencies of individual elements are again determined by formula (\ref{a6_01}), but $n_j$ in this case represents the $j$-th element of the ordered sequence of integers $n_j\in {1,2,\ldots,N}$.
\end{itemize}
Note that when deriving the formula for the power spectral density (\ref{a4_21}), we did not specify in advance the specific algorithm for sampling the natural frequencies. Within the linear approximation, the obtained formula remains valid for both regular and random sampling. However, in \cite{Kirillov2025} it was noted (without detailed justification) that it is preferable to choose a random rather than regular frequency distribution. Below we analyze the reasons for this and discuss the mechanisms leading to significant discrepancies between theory and numerical experiment for regular distribution.

\begin{figure*}
\center{\includegraphics[width=14cm]{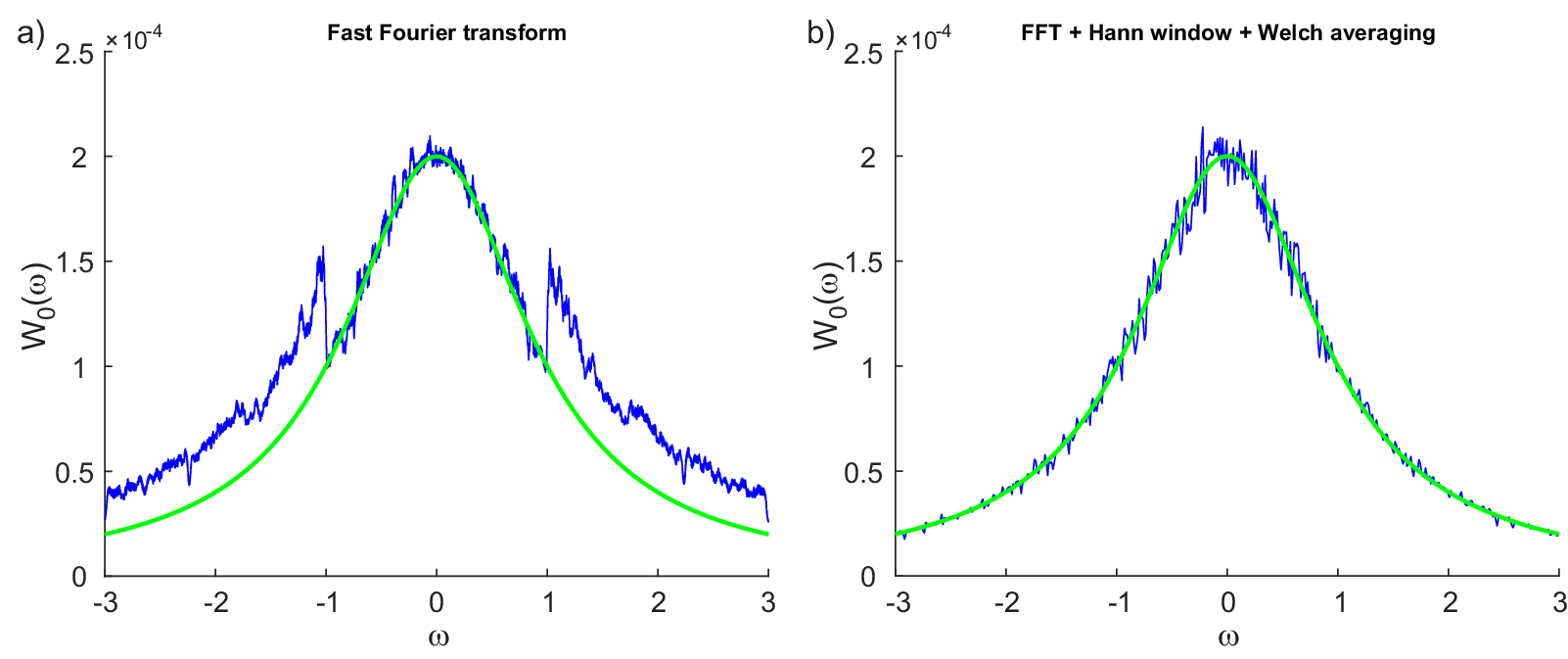}}
\caption{Comparison of the power spectral density obtained from the analytical formula (\ref{a3_05}) and constructed numerically: \textit{a} - using classical fast Fourier transform with moving average window $\Delta\omega_{aver}=0.03$; \textit{b} - using Welch's method with a Hann window and 50\% overlap of segments of length $2^{23}$. Parameters: $N=10^4$, $T=2\cdot 10^4$, discretization step $\Delta t=1 \cdot 10^{-4}$.} \label{fig:Spctr_Welch}
\end{figure*}

First, let us make a brief remark about the features of the discrete Fourier transform. The reasons for this can be most clearly demonstrated by revisiting the case where the connections between system elements are absent. As shown earlier, the output signal of such a system is a simple sum of individual harmonics, and its spectrum can be easily represented in strict analytical form. Figure~\ref{fig:Spctr_Welch}, \textit{a} compares the corresponding theoretical curve with the shape of the power spectral density of the finite-dimensional system constructed using the fast Fourier transform.
It can be seen that while the theoretical spectrum changes monotonically on both sides of the center frequency, the spectrum of the finite-dimensional system contains two clearly distinguishable jumps, resulting in the overall picture appearing as a patchwork of several fragments. In the central region, good qualitative and quantitative agreement between the theoretical and numerical graphs is observed. Meanwhile, the edges of the numerically constructed spectrum are noticeably higher than the theoretical curve. This behavior is related, on one hand, to the discrete nature of the finite-dimensional system spectrum, and on the other hand, to the features of the Fourier transform for finite-duration signals.
The spectrum $W_{\mathcal{F}}(\omega)$ obtained by this transform represents the convolution of the true spectrum of the infinite signal $W(\omega)$ with the Fourier transform of the rectangular window $\Pi(\omega)$:
\begin{equation}\label{a7_01}
W_{\mathcal{F}}(\omega)=W(\omega)\ast \Pi(\omega),
\end{equation}
where
\begin{equation}\label{a7_02}
\Pi(\omega)=T\sinc \bigg(\frac{\omega T}{2\pi}\bigg)=\frac{\sin (\omega T / 2)}{\omega / 2},
\end{equation}
and the parameter $T$ determines the signal duration.

The distortions in the spectrum $W_{\mathcal{F}}(\omega)$ arise due to the influence of the side lobes of the function $\Pi(\omega)$, which describes the Fourier transform response to a single spectral component. A key role is played by interference effects that occur when contributions from the convolution of individual spectral components $W(\omega)$ with the kernel $\Pi(\omega)$ overlap. Since the density of spectral components is non-uniform in frequency, the nature of interference changes significantly when moving from the peripheral regions of the spectrum to the center.

As the signal duration $T$ increases, the width of the main lobe of $\Pi(\omega)$ decreases, leading to increased frequency resolution. The width of the side lobes also decreases. However, their amplitude decays relatively slowly, following a $\sim 1/\omega$ law. This ensures a large extent of the region influenced by the side lobes, potentially covering the entire frequency range.

The nature of the interference pattern evolves when moving from the periphery of the spectrum to its center. At the periphery, the density of natural frequencies is low, individual harmonics are sparsely located, and the spectrum $W_{\mathcal{F}}(\omega)$ in their vicinity reproduces the shape of the $\sinc$ function with a pronounced main peak and decaying side lobes. In this region, interference is practically absent, and each spectral component makes an independent contribution determined by the convolution with $\Pi(\omega)$.

As one moves toward the central region of the spectrum, the density of harmonics increases, individual $\sinc$ functions begin to overlap, and their lobes start to interfere. In the spectrum, regions of constructive interference appear, where the lobes predominantly add up, as well as regions of destructive interference, where the lobes partially cancel each other.

In the central region of the spectrum, the density of harmonics is maximal. The function $\Pi(\omega)$ in this case overlaps with many spectral components simultaneously, and the results of their convolutions overlap with each other. As a result, the effects of constructive and destructive interference compensate each other. This ensures that the numerically constructed spectrum $W_{\mathcal{F}}(\omega)$ is close to the theoretical distribution.

The region of applicability of the fast Fourier transform can be quantitatively characterized by the boundary frequencies $\pm\Delta\omega$. They define the interval beyond which the level of spectral distortion becomes significant. The width of this interval is inversely proportional to the signal duration $\sim 1/T$. Thus, contrary to natural expectations, as the signal duration increases, the region of correct spectral reproduction narrows, and for sufficiently large $T$, spectral distortions can cover almost the entire frequency range, including the central frequency.

When frequencies are chosen randomly, the positions of individual harmonics $W(\omega)$ become statistically independent, disrupting the phase relationships between the contributions of the lobes of different spectral components to $W_{\mathcal{F}}(\omega)$. As a result, the effects of constructive interference are suppressed even in the transition region between the center of the spectrum and its periphery, and the observed distortions are significantly weakened, becoming practically indistinguishable against the theoretical spectrum.

In the following, when constructing the spectrum, we will use the Hann window instead of the rectangular window. Its use is a classic method for suppressing side lobes, ensuring their decay $\sim 1/\omega^3$. Additionally, to reduce the variance of the power spectral density estimate, we will use Welch's averaging method with 50\% overlap. The numerically constructed spectrum is shown in Fig.~\ref{fig:Spctr_Welch}, \textit{b}. It can be seen that unlike Fig.~\ref{fig:Spctr_Welch}, \textit{a}, the analytical and numerically constructed curves almost coincide. This improvement in spectral representation is achieved at the cost of reduced frequency resolution. We note again that the appearance of undesirable effects when using direct discrete Fourier transform is most pronounced when using a regular frequency distribution, while for a random distribution their influence is almost imperceptible. This is one of the reasons why random frequency distribution is preferable.

\begin{figure*}
\center{\includegraphics[width=14cm]{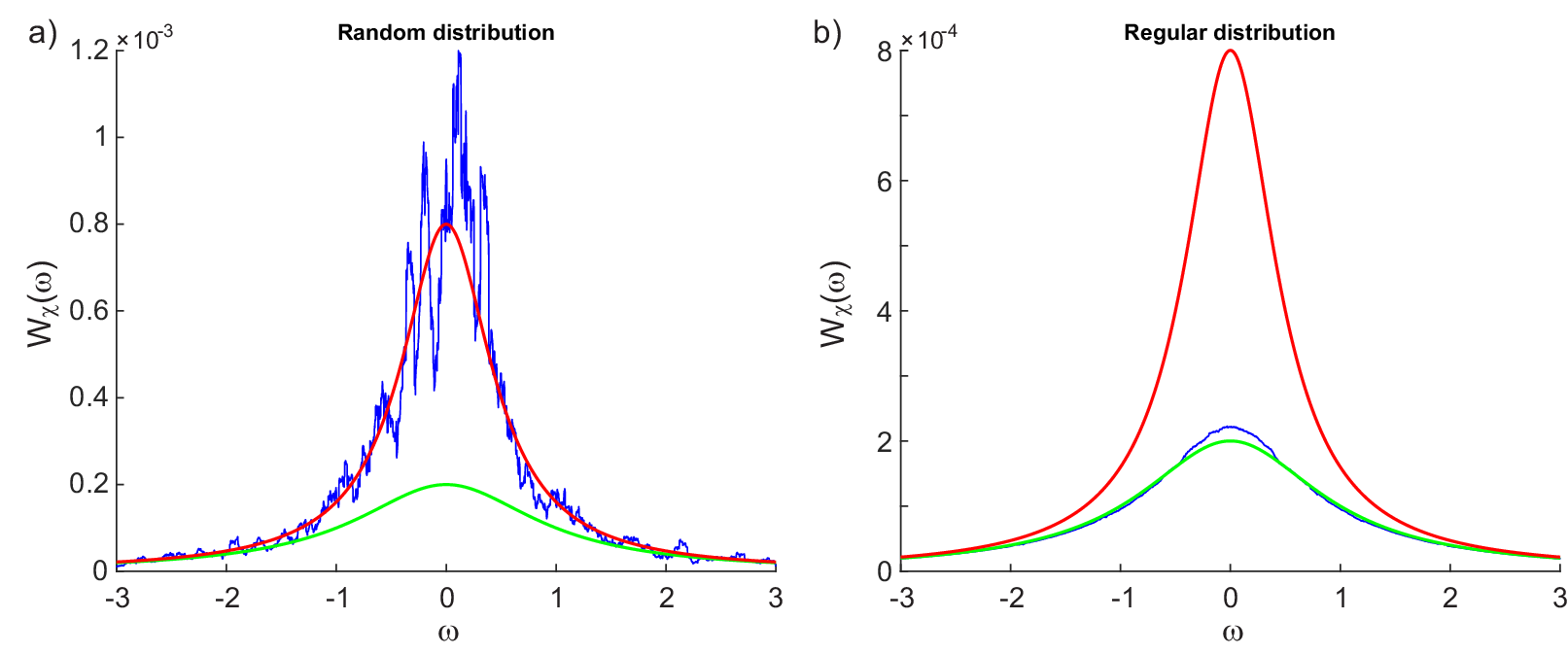}}
\caption{Comparison of the power spectral density obtained from the analytical formula (\ref{a4_21}) and constructed numerically for \textit{a} - random and \textit{b} - regular distribution of natural frequencies. Red curve: theoretical shape of the shot noise spectrum; blue: numerical simulation results; green: theoretical spectrum of free shot noise.} \label{fig:Spctr_compare_1}
\end{figure*}

Let us now proceed directly to analyzing the properties of oscillations in the presence of inter-element interaction. Fig.~\ref{fig:Spctr_compare_1}, \textit{a,b} present graphs of the power spectral density of shot noise obtained from numerical integration for $N=10^4$ and for both frequency selection methods at fixed parameter values $K=1$, $\omega_0=0$, $\Delta=1$. The system integration was performed for a duration of $T=2\times 10^5$ using the Euler method with a step $\Delta t=1/N=1\cdot 10^{-4}$ (inversely proportional to the network size, which is due to the shape of the natural frequency distribution), with the interval $t\in[0;T/2]$ excluded from the spectrum calculation to allow transients to decay. The initial phase values were chosen randomly and independently from a uniform distribution on the interval $\varphi\in[0;2\pi)$. The power spectrum of the Kuramoto order parameter \eqref{a1_02} was calculated using the fast Fourier transform with a Hann window. Spectral analysis included Welch averaging with a segment length of $2^{26}$, as well as additional averaging of the obtained spectrum with a moving window of width $\Delta\omega=7.5\cdot 10^{-2}$.

The numerical simulation results are compared in Fig.~\ref{fig:Spctr_compare_1}, \textit{a,b} with the theoretical estimate \eqref{a4_21} and the free shot noise spectrum (\ref{a3_05}). It can be seen that for random frequency selection, despite the presence of local spikes and dips in the spectrum, the numerical results on average agree well with the theoretical estimate. For deterministic frequency selection, a significant discrepancy between the numerical and theoretical results is observed: the height of the main peak differs by a factor of several from the theoretical predictions.

\begin{figure*}
\center{\includegraphics[width=12cm]{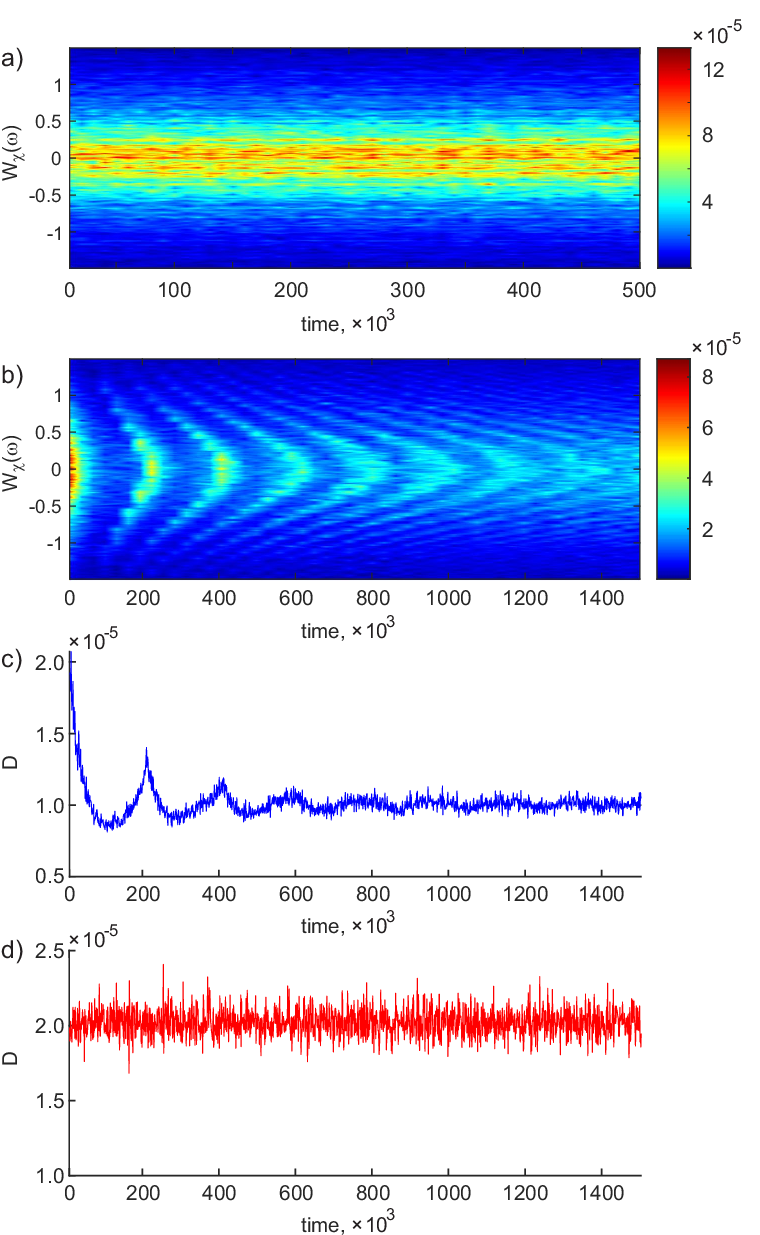}}
\caption{Time-frequency diagrams of the evolution of the power spectral density of shot noise \( W_{\chi}(\omega, t) \) for a network of \( N = 10^5 \) oscillators at \( K = 1 \). Color encodes the intensity of the power spectral density. \textit{a} - Random frequency selection: the spectrum shape is stable over time, only local fluctuations due to finite network size are observed; \textit{b} - deterministic (quasi-uniform) frequency selection: spectral peaks drift, generating a wave-like structure. Dynamics of the order parameter variance \( D(t) \) calculated over moving windows: \textit{c} - for deterministic frequency selection (blue curve), periods of increased and decreased synchronization are observed, correlating with the wave-like changes in the spectrum; \textit{d} - for random selection (red curve), no such modulation is present.}
\label{fig:Freq_time_1}
\end{figure*}

To identify the reasons for the discrepancy, we conducted a series of numerical experiments with different numbers of elements in the network (from $N=10^3$ to $N=10^5$). The analysis showed that in the absence of inter-element interactions, the numerically constructed spectra agree with the theoretical free shot noise spectrum both qualitatively and quantitatively (see Fig.~\ref{fig:Spctr_Welch}, \textit{b}). In this case, the phases of the network elements $\varphi$ at each moment are independent and uniformly distributed on the interval $\varphi_j\in[0;2\pi)$. Our assumption, confirmed numerically, was that in the case of weak interactions (which includes the case $K=1$ in the Kuramoto model), collective effects accumulate slowly. When starting from statistically identical initial conditions (uniform phase distribution) and statistically identical distributions of natural frequencies, differences in the power spectral density of the order parameters build up gradually. This allows the use of windowed Fourier transform to analyze the dynamics of the spectra. For this purpose, the entire time interval over which numerical integration was performed was divided into short segments of length $\Delta T$ (depending on $N$). For each such segment, the power spectrum of the signal $R(t)$ was calculated (using the Hann window and Welch averaging). Fig.~\ref{fig:Freq_time_1}, \textit{a,b} present diagrams of the evolution of the collective fluctuation spectra for a network of $N=10^5$ elements. Significant differences are observed in the behavior of the spectra for the cases of random and deterministic frequency selection.

\begin{figure*}
\center{\includegraphics[width=12cm]{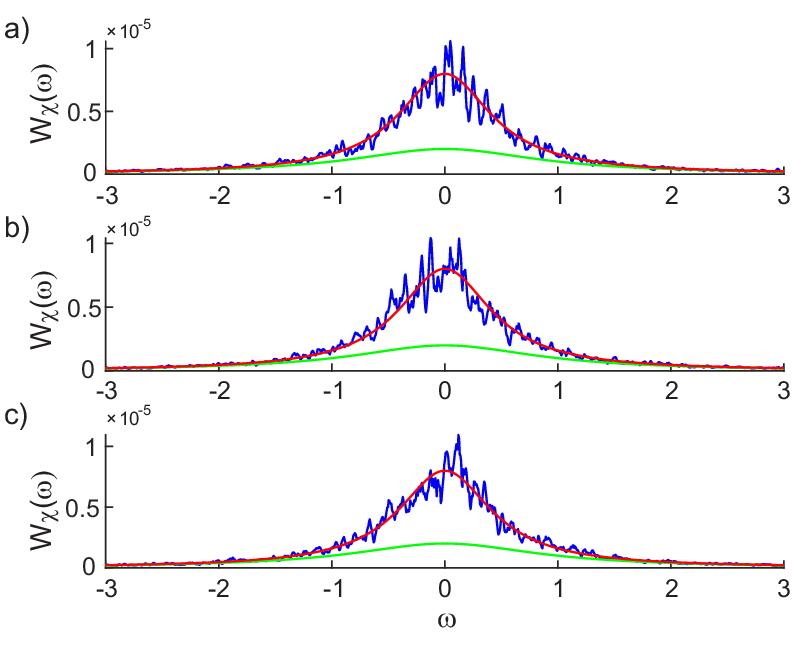}}
\caption{Local power spectral density for random natural frequency distribution for three time intervals: \textit{a} - $t\in[0,10\cdot 10^3]$, \textit{b} - $t\in[490\cdot 10^3, 500\cdot 10^3]$,  \textit{c} - $t\in[1490\cdot 10^3,1500\cdot 10^3]$. Red curve: theoretical shape of the shot noise spectrum; blue: numerical simulation results; green: theoretical spectrum of free shot noise.} \label{fig:Spctr_compare_mult_aver_rand_v3}
\end{figure*}

With random frequency distribution, the spectrum shape remains stable over time and is on average well approximated by the theoretical curve.
This is illustrated in Fig.~\ref{fig:Spctr_compare_mult_aver_rand_v3}, \textit{a-c}, which shows local spectra for three different time intervals of equal duration. At the same time, local spikes and dips are observed in the constructed spectrum, which may appear and disappear over individual time intervals, but their overall position remains preserved. Such individual spectral features are due to the corresponding local maxima and minima of the natural frequency distribution and are related to the finite size of the network.

\begin{figure*}
\center{\includegraphics[width=12cm]{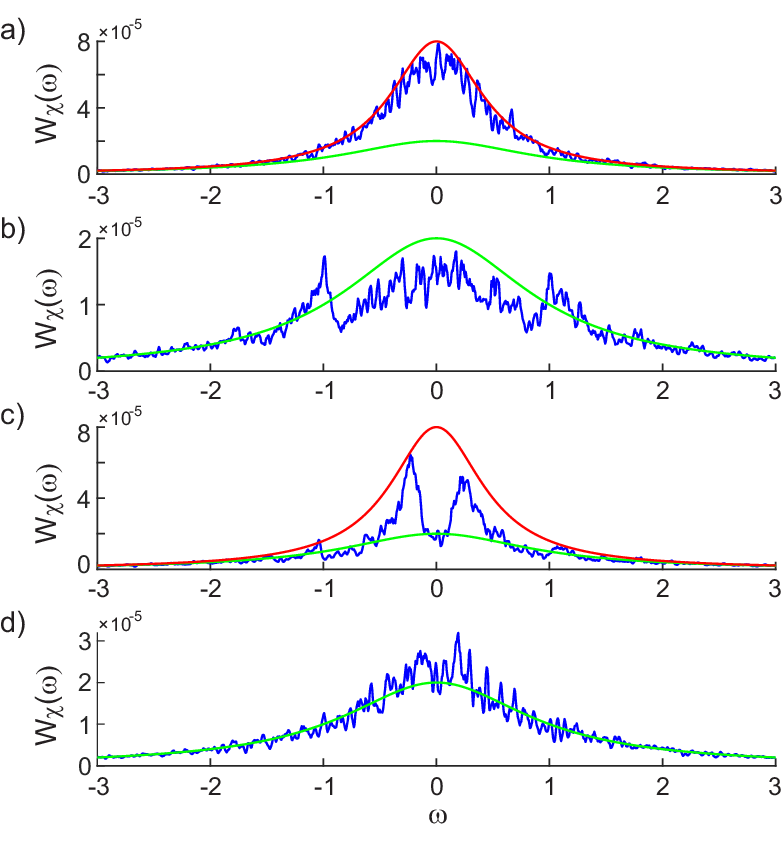}}
\caption{Spectra at different times. \textit{a} - $t\in[0,10\cdot 10^3]$, \textit{b} - $t\in[100\cdot 10^3, 110\cdot 10^3]$,  \textit{c} - $t\in[190\cdot 10^3, 200\cdot 10^3]$,  \textit{d} - $t\in[1490\cdot 10^3,1500\cdot 10^3]$. Red curve: theoretical spectrum; green: free shot noise spectrum; blue: spectrum of the microscopic network.} \label{fig:Spctr_compare_mult_aver_v2}
\end{figure*}

In the case of regular frequency distribution, the spectral dynamics change fundamentally. While in the initial stage the spectrum is close to the theoretical one (see Fig.~\ref{fig:Spctr_compare_mult_aver_v2}, \textit{a}), over time it undergoes significant qualitative and quantitative changes. Over time, the central maximum in the spectrum begins to sag (see Fig.~\ref{fig:Spctr_compare_mult_aver_v2}, \textit{b}). Simultaneously, local maxima form on the left and right wings of the spectrum, which move toward the center, where they meet and cancel each other (see Fig.~\ref{fig:Spctr_compare_mult_aver_v2}, \textit{c}). New local maxima then appear on the spectrum wings, and the process repeats. As a result, the spectrum exhibits slow oscillations, during which the height of the central peak changes by a factor of several, and a wave-like structure forms on the time-frequency diagram (see Fig.~\ref{fig:Freq_time_1}, \textit{b}). The characteristic period of the slow spectral oscillations is on the order of $T_{osc}\sim 202\cdot 10^3$, with oscillations decaying over a characteristic time $T_{fade}\sim 1500\cdot 10^3$ (sufficient for transients to decay). After the transients have subsided, the system reaches a new dynamically equilibrium state that falls outside the theoretical description. Remarkably, as shown in Fig.~\ref{fig:Spctr_compare_mult_aver_v2}, \textit{d}, in this state the spectrum shape turns out to be close to the theoretical estimate (\ref{a3_05}) for free shot noise.

To better understand the nature of the emergence of new local maxima in the deterministic case, let us examine the system dynamics in more detail. Since the distribution of natural frequencies of the oscillatory elements according to (\ref{a2_09}) has a bell shape, most of them are concentrated near the central frequency. With a regular internal structure of the distribution, the frequencies of neighboring elements in this region change with an almost constant step. This creates favorable conditions for resonant interactions between oscillatory modes.

The highest intensity of resonant processes occurs precisely in the central part of the spectrum due to the high density of oscillators. This leads to active energy transfer between modes in this region and, as a consequence, to a decrease in the power spectral density at the central frequency and its redistribution to the peripheral regions of the spectrum. As one moves away from the spectrum center, the step size in frequency between neighboring elements rapidly increases. This hinders further energy transfer to the spectrum periphery. As a result, energy begins to concentrate on a limited number of modes, causing their amplitude to increase and forming pronounced local peaks in the power spectral density.

Over time, the position of these peaks on the frequency axis slowly shifts toward the central frequency. Based on general physical considerations, it can be assumed that this drift is due, on the one hand, to weak nonlinear effects causing fluctuations in the natural frequencies of the oscillators, and on the other hand, to the system's tendency toward dynamic equilibrium. Modes at the periphery, having received excess energy, tend to return it to the central modes.

The described process is oscillatory in nature. As one spectral peak begins to drift from the periphery toward the center, the next peak appears in its place, formed due to the continuing energy flow from the central zone. As a result, a sequential chain of localized peaks of the power spectral density propagates toward the center. However, over time, this oscillatory process gradually decays, and the system reaches a new state of dynamic equilibrium in which the energy flows from the center to the periphery and back balance each other. Consequently, the power spectral density stabilizes, and its shape no longer undergoes significant changes.

In contrast to the regular distribution, with random frequency distribution resonant conditions are rare and random. In this case, interactions between elements are incoherent and non-resonant, and the influence of nonlinearity is significantly weakened. As a result, energy transfers between modes are minimal, and the power spectral density remains on average stable, experiencing only some fluctuations.

The slow oscillations of the shot noise spectrum shape, primarily affecting the height of the central peak, indicate a slow change in the degree of partial synchronization of the system, which is confirmed by the dynamics of the variance $D(t)$ of the Kuramoto order parameter, illustrated in Fig.~\ref{fig:Freq_time_1}, \textit{c}. Here, the total signal duration $T$ is divided into a sequence of intervals $\Delta T=10^3$, and the corresponding variance value is calculated for each interval. The graph clearly shows periods of increased and decreased synchronization, coinciding in duration with the slow oscillations of the shot noise spectrum. Note that such transient synchronization is observed only for deterministic frequency selection, while for random selection no transient synchronization is observed.

\begin{figure*}
\center{\includegraphics[width=14cm]{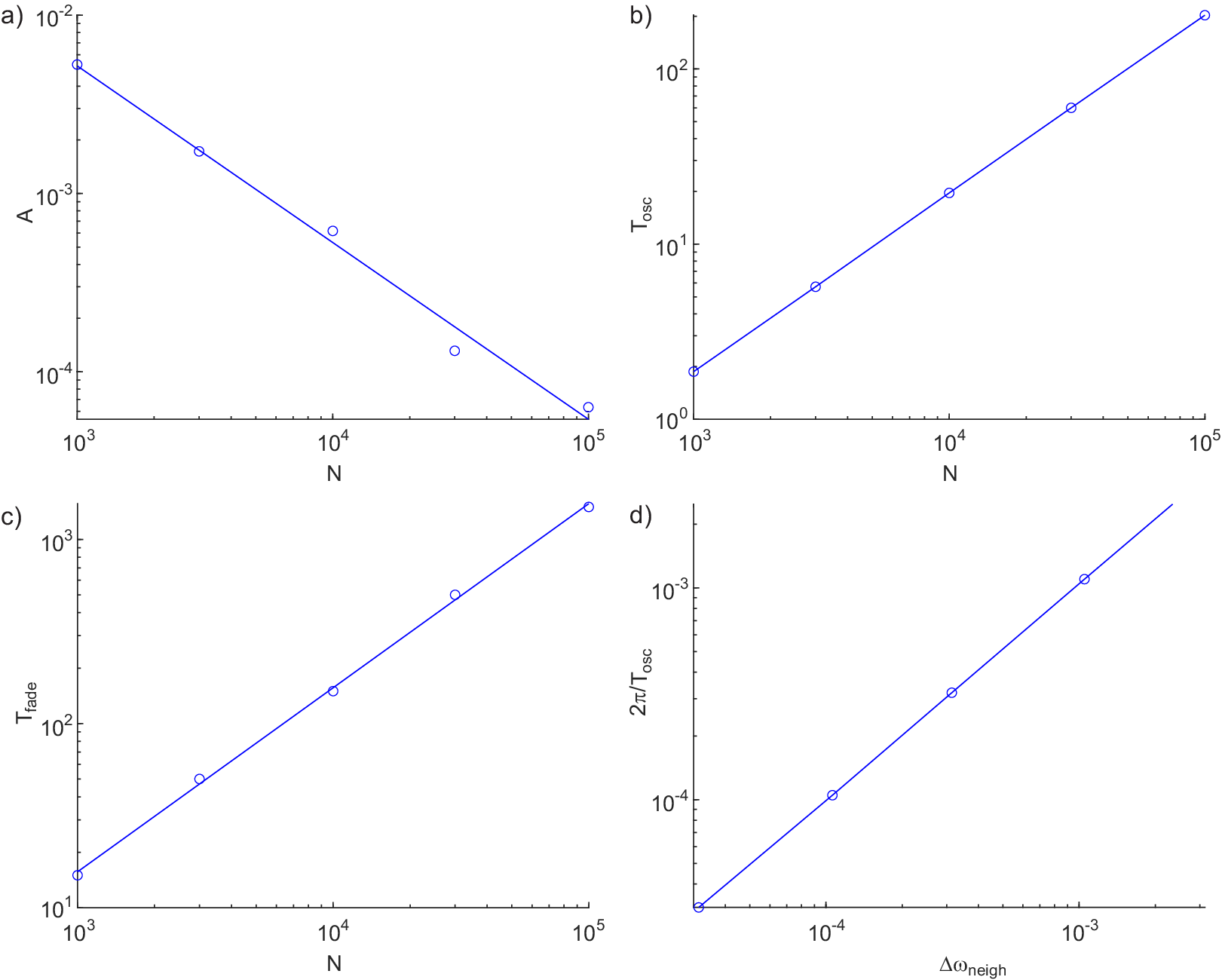}}
\caption{\textit{a} - Amplitude (range) of power spectral density oscillations as a function of the number of network elements; \textit{b} - Period of spectral oscillations; \textit{c} - Decay time of spectral oscillations. \textit{d} - Relationship between the period of slow oscillations and the difference between the frequencies of neighboring elements selected deterministically near the peak of the natural frequency distribution.} \label{fig:Lin_scale}
\end{figure*}

We investigated the dependence of the slow spectral oscillations on the network size $N$, limiting ourselves to five values $N: 10^3, 3\cdot 10^3,10^4,3\cdot10^4,10^5$ due to the computational cost of numerical experiments. For each network size, the following quantities were estimated:
\begin{itemize}
    \item amplitude of spectral oscillations as the difference between the maximum and minimum observed heights of the peaks of local spectra;
    \item period of spectral oscillations as the time between two adjacent local maxima of the peak height of local spectra;
    \item decay time of spectral oscillations, as the time after which the transient process can be considered complete.
\end{itemize}
The obtained estimates are shown in Fig.~\ref{fig:Lin_scale}, \textit{a-c}. 
The linear scaling of the dynamic process parameters (period, amplitude, decay time) as the number of network elements changes by two orders of magnitude serves as confirmation of the robustness (``coarseness'') of the described regime. This indicates that (according to Fig.~\ref{fig:Lin_scale}, \textit{b}) the resonant conditions formed by the regular structure in the natural frequency spectrum are preserved over a wide range of network sizes, and nonlinearity does not lead to the emergence of new dynamic effects. Most interesting is the observation that the period of slow oscillations $T_{osc} \sim N$, which corresponds to a frequency $\omega_{osc}\sim N^{-1}$. Note that (as confirmed by the data in Fig.~\ref{fig:Lin_scale}, \textit{d}) this frequency corresponds to the difference between the frequencies of neighboring elements $\omega_{osc}=\Delta\omega_{neigh}$, selected deterministically from distribution \eqref{a2_09} near its peak.

\section{Discussion and conclusions}

In this work, we investigated the influence of the microscopic realization of the natural frequency distribution on the collective dynamics of finite ensembles of Kuramoto phase oscillators in the subcritical regime (\( K < 2 \)). Using the previously developed approach based on the concept of shot noise and the nestling principle, we obtained an analytical expression for the power spectral density of collective fluctuations (38). This expression is in good quantitative agreement with the results of numerical simulations for the case of random sampling of frequencies from a Lorentzian distribution.

However, the main result of this work is the discovery of a qualitatively new effect that arises when the natural frequencies of the oscillators are chosen deterministically (quasi-uniformly). In this case, despite the identity of the integral distribution function \( g(\omega) \), the collective behavior of the system is fundamentally different. Anomalously slow oscillations are observed in the shot noise spectrum, manifested as ``breathing'' of the shape of the local spectrum calculated over finite time intervals. These oscillations primarily affect the amplitude of the central peak, which corresponds to a slow change in the degree of partial synchronization in the system.

A key observation is that the characteristic period of the observed oscillations \( T_{\text{osc}} \) scales linearly with the system size \( N \), and the corresponding frequency \( \omega_{\text{osc}} \sim N^{-1} \). This frequency coincides with high precision with the spacing between neighboring natural frequencies of oscillators near the distribution center (where the frequency of occurrence of elements is maximal). For random sampling, the frequencies are irregularly located, and such a distinct frequency difference does not occur. In the deterministic case, however, the frequencies form an ordered structure, and it is likely that the presence of this regularity in the microscopic distribution introduces a new, slow time scale into the system.

Our results indicate that the identity of the integral form of the distribution of parameters of individual elements in a population does not guarantee the equivalence of their collective dynamics. Regularity at the microscopic level, even when hidden behind a smooth macroscopic density, can initiate complex transient processes and qualitatively change the spectral properties of the system. The discovered effect of slow spectral oscillations demonstrates the limitations of purely mean-field descriptions that operate only with the probability density function \( g(\omega) \). It highlights the necessity of accounting for the fine structure of the distribution for correctly predicting the behavior of finite ensembles, especially in problems where the temporal evolution of the system, rather than just its stationary characteristics, is important. The results obtained lay the foundation for constructing generalized theoretical models that account for the influence of such hidden correlations on collective dynamics.

\begin{acknowledgments}
The theoretical part of the work was supported by the Ministry of Science and Higher Education of the Russian Federation under the state assignment for the Institute of Applied Physics RAS, project FFUF-2024-0011, and the numerical simulations were supported by the Russian Science Foundation grant No. 25-22-00660.
\end{acknowledgments}

\section*{Data Availability Statement}

The data that support the findings of this study are available from the corresponding author upon reasonable request.

\section*{references}

\end{document}